\documentclass[a4paper]{jpconf}
\usepackage{graphicx}
\usepackage{amssymb}        
\usepackage{amsmath}
\newcommand{\be}{\begin{equation}}
\newcommand{\ee}{\end{equation}}
\newcommand{\beq}{\begin{eqnarray}}
\newcommand{\eeq}{\end{eqnarray}}

\newcommand{\vect}[1]{{\bf #1}}

\newcommand{\diff}{{\mathrm d}}

\begin{document}
\title{On the Seismicity of September 7, 2011 X1.8-class Flare}

\author{{S.~Zharkov}$^{1,2}$, {L.M.~Green}$^2$, {S.A.~Matthews}$^2$, {V.V.~Zharkova}$^3$}

\address{$^1$ Department of Physics and Mathematics, University of Hull, Cottingham Road, Kingston-upon-Hull, HU6~7RX, UK}
\address{$^2$ Mullard Space Science Laboratory,
          University College London, Holmbury St. Mary, Dorking, RH5 6NT,  UK}
\address{$^3$ Department of Mathematics, University of Bradford, Bradford, {BD7 1DP}, UK}

\ead{s.zharkov@hull.ac.uk}

\begin{abstract}
We present results of our preliminary analysis of acoustically active X-class flare of September 7, 2011. We report two  acoustic sources detected via acoustic holography and verified by finding a ridge in time-distance diagrams. We compare the directional information extracted from time-distance and acoustic holography, showing a good agreement in this case. We report that the direction where amplitude of the wave-front is the largest lies through the strong magnetic field and sunspot, suggesting that absorption of the acoustic wave power by magnetic field can be ruled out as a wave anisotropy mechanism in this case. 
\end{abstract}

\section{Introduction}
Although theoretically predicted \cite{wolff72} flare driven sun-quakes have only been discovered relatively recently \cite{kz1998} using SOHO/MDI data. Detection and analysis of sunquakes in the previous solar cycle 23 \cite[and others]{DL2005,K2006,Moradi2007,Martinez2008a} has been considerably restricted by the availability of suitable observations, see recent review \cite{donea11} for more details. Recent launch of SDO/HMI \cite{ScherrerHMI2011} combined with the approaching maximum of the current Solar Cycle 24 now ensures that we can observe this phenomenon in new unprecedented detail due to better data coverage as well as improved spatial and temporal resolution.

Sunquakes, seen as near circular ripples, propagating outward from impulsive solar flares along the solar surface, appear 20\,-\,60 minutes after the flare onsets. Strong anisotropy of the surface wave-field has often been reported \cite{donea11,K2006,Kosovichev2007}.  This has normally been attributed to the suppression and absorption of acoustic waves by magnetic field, which has been extensively studied theoretically and modelled \cite{Moradi2010}. It has also been observed that, in sunquakes, the largest amplitudes  of the generated surface wavefront are in the direction of the apparent moving source.

\begin{figure}
\begin{center}
\begin{minipage}[b]{5.cm}
\includegraphics[width=4.75cm]{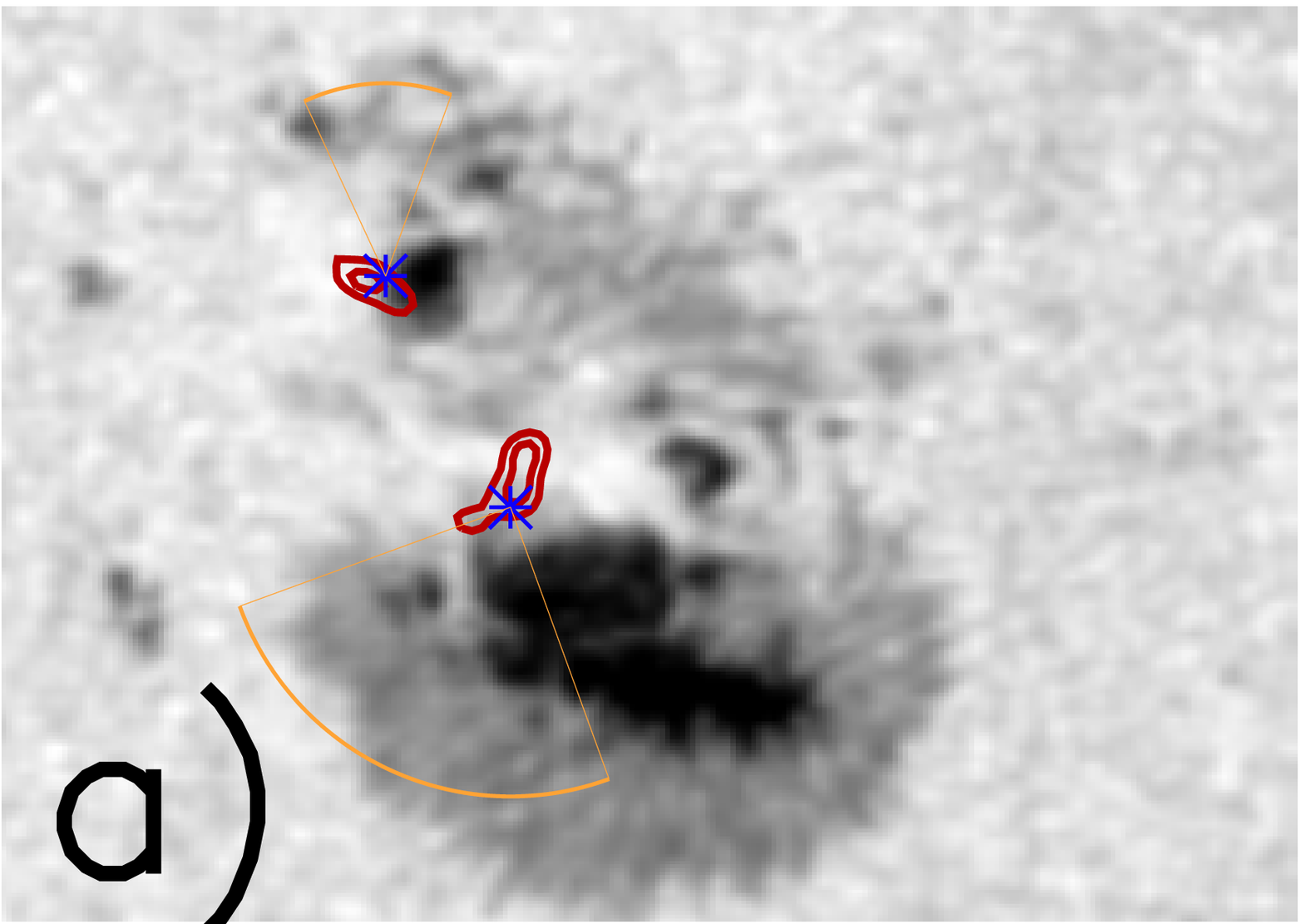}
\includegraphics[width=4.75cm]{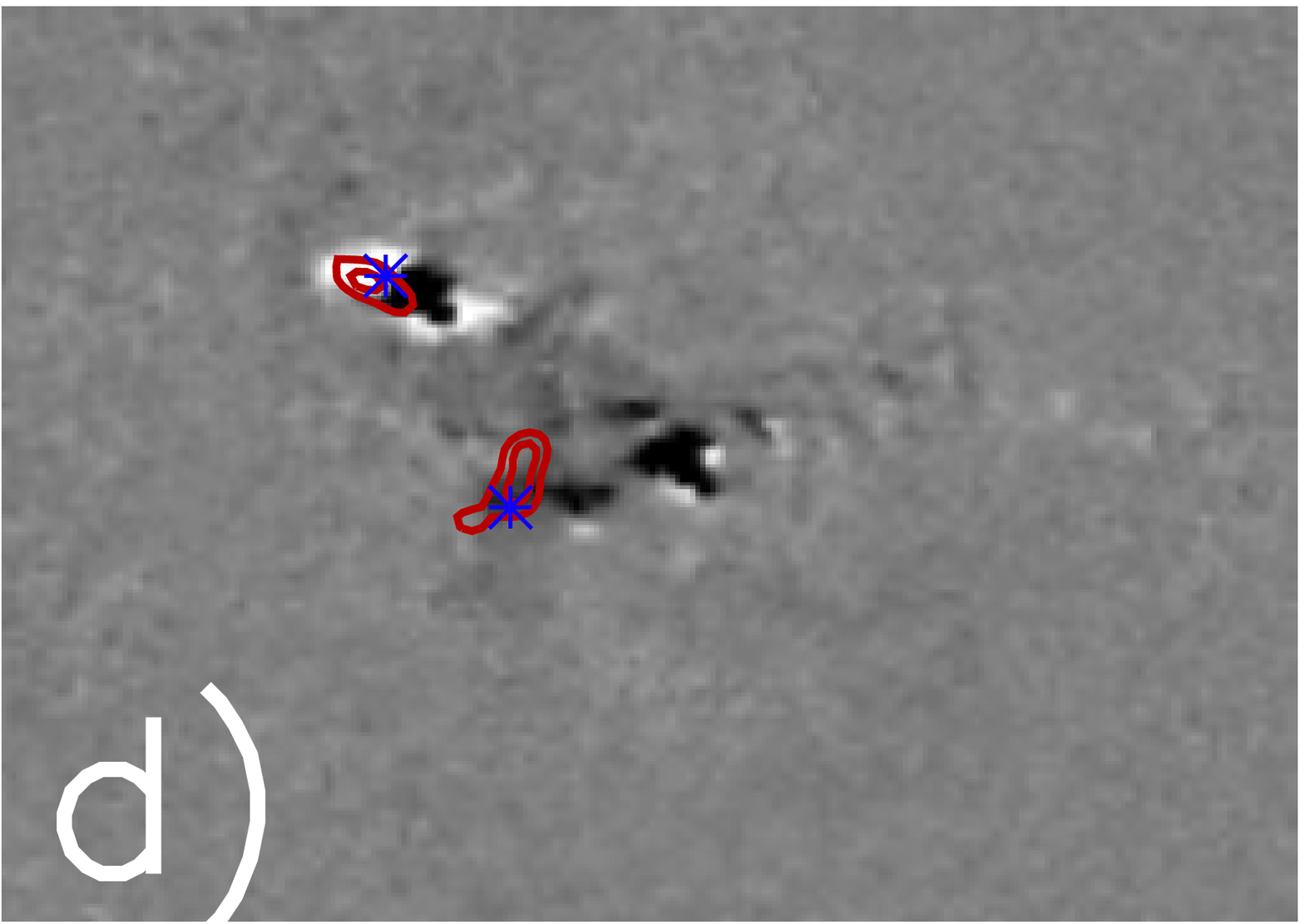} 
\includegraphics[width=4.75cm]{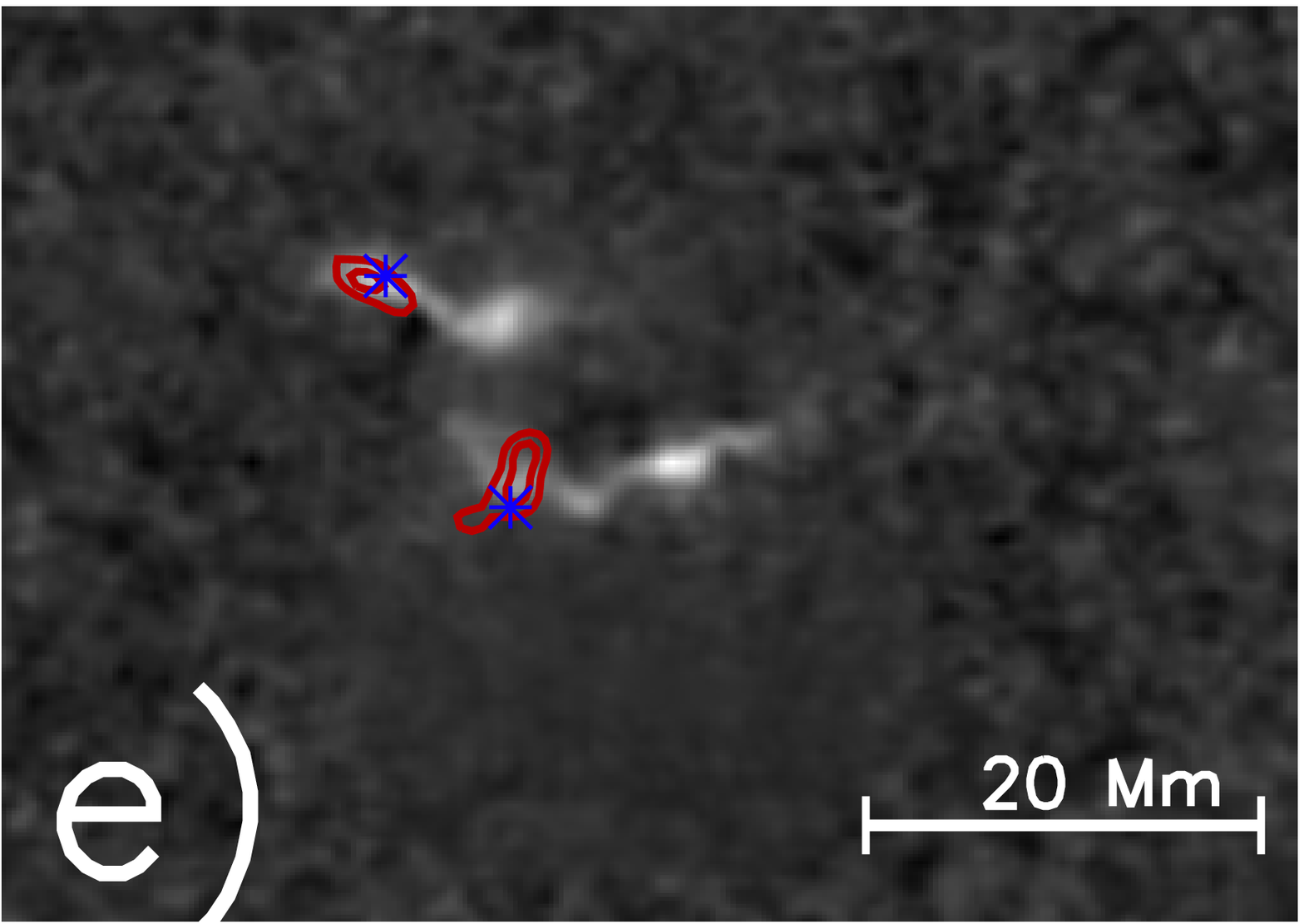}
\end{minipage}
\begin{minipage}[b]{10.5cm}
\begin{center}
\includegraphics[width=4.75cm]{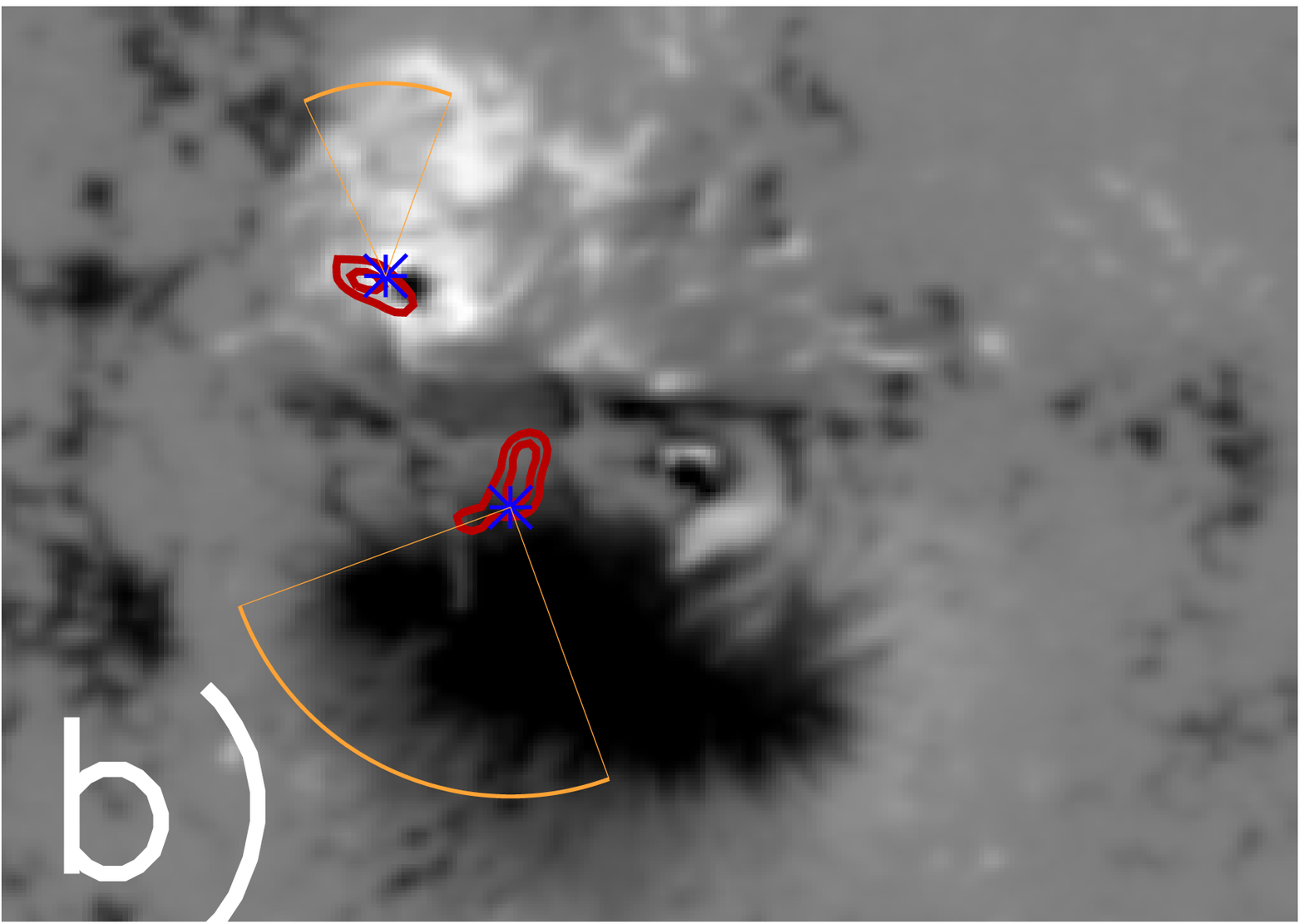}
\includegraphics[width=4.75cm]{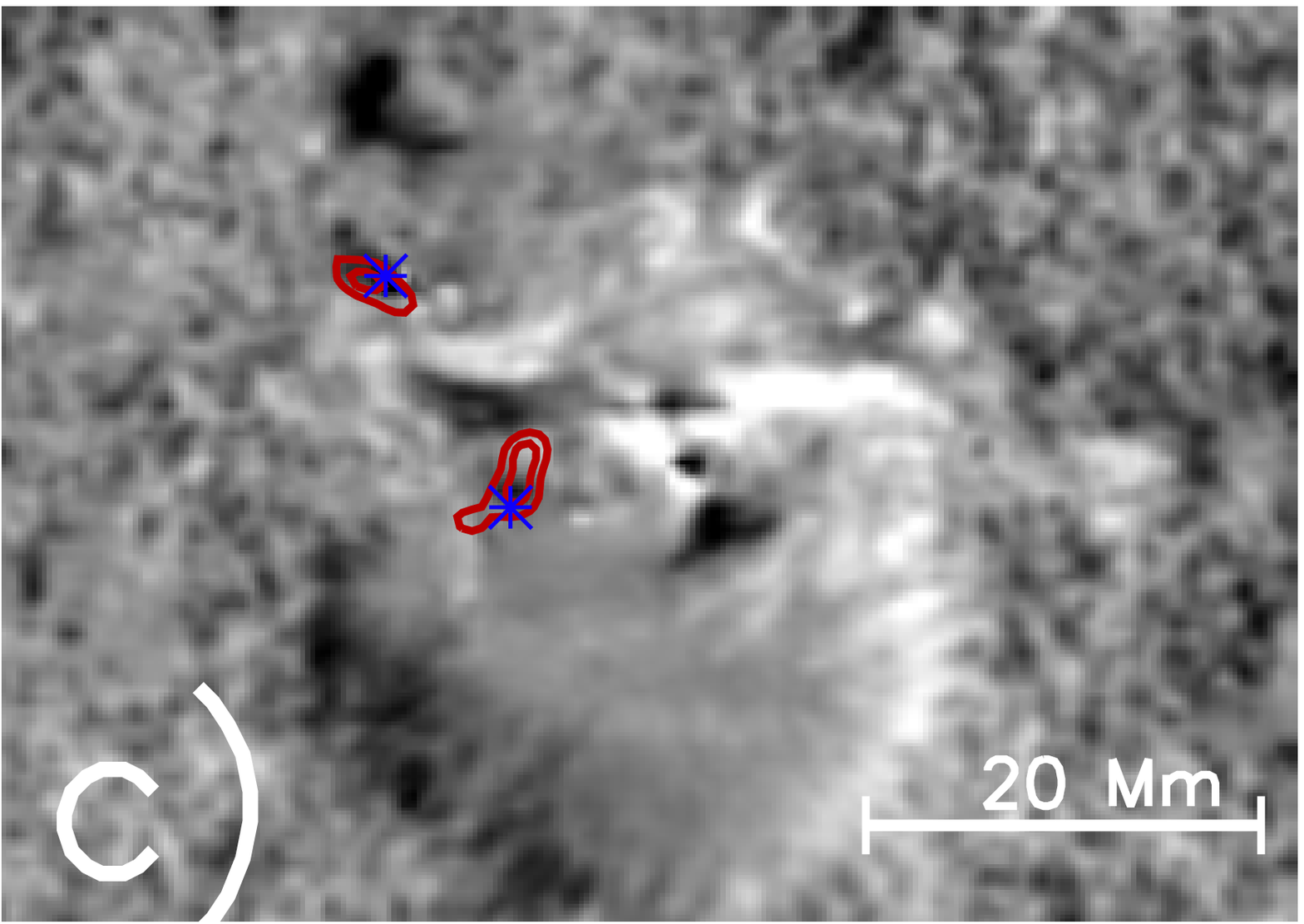}
\includegraphics[width=10.2cm, height=6.7cm]{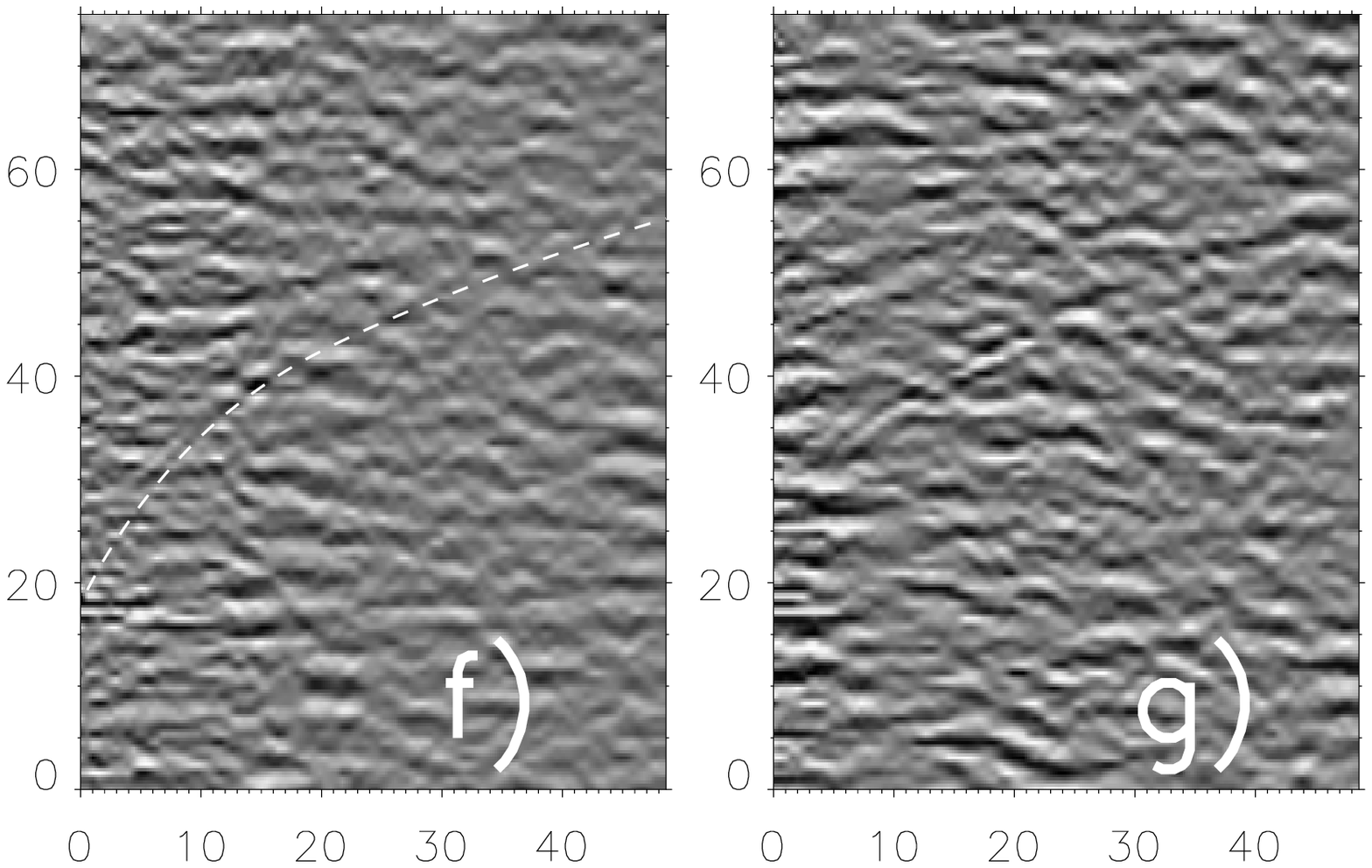}
\end{center}
\end{minipage}
\end{center}
\caption{\label{fig:2sources} Context images taken at 22:37:22\,UT: a) HMI continuum intensity saturated (black) at 30\% of quiet sun intensity,
b) line-of-sight magnetogram, scaled between $\pm1$ kG and c) line-of-sight doppler velocity on the scale between $\pm1.5 {\rm \,km\,s}^{-1}$. 
The running difference images taken around 22:36-22:37\,UT are shown in d) for HMI magnetogram (scaled between $\pm200$ Gauss)
 and e) continuum showing variation between $-4\%$ and $20\%$ of quiet sun intensity. 
 The red contours represent the peak 6 mHz egression power at 2 and 3 factor the quiet Sun emission ($\approx 288 {\rm \, m^2\,s^{-2}}$). Blue stars indicate the locations of found time-distance ridges, while orange arcs on images at the top row indicate the direction of the detected time-distance ridges. The FOV and scale are the same across a)-e).
The time-distance diagrams computed for the two sources are presented in panels f) for the northern Source 1 with theoretical time-distance curve overplotted in white and g) for the southern Source 2. In time-distance plots, distance in Mm is plotted along $x$-axis, time in minutes since 22:20:07\,UT is plotted along $y$-axis. }
\end{figure}

\section{Data and Method} \label{sec:method} 
 GOES Soft X-ray flare list reports the X1.8 class flare  taking place in NOAA Active Region 11283 on September 7, 2011, starting at 22:32\,UT and peaking at 22:38\,UT. The flare was accompanied by a CME as reported by SOHO/LASCO and CaCTUS CME event catalogues. 

We use full disk SDO/HMI intensity, dopplergram and line-of-sight magnetic field data at 45-second cadence to produce three hour-long datacubes, which are extracted by remapping and derotating the region of interest using Postel projection and the Snodgrass differential rotation rate. The spatial resolution of the remapped data is 0.04 degrees per pixel. We apply acoustic holography to calculate the egression power maps from observations \cite{BL1999, Donea1999, LB2000} along with computing time-distance diagram \cite{kz1998,K2006,Kosovichev2007,Zharkova07}.  The egression power is computed for each integral frequency from 3 to 10 mHz, by applying 1-mHz frequency bandwidth filters to the data and using Green functions built for surface monochromatic point source of corresponding frequency using geometrical optics approach. The choice of 1 mHz, $\Delta \nu$, bandwidth implies timing uncertainty of $\Delta t = 1/ \Delta \nu=1/1000$ seconds \cite{DL2005}. The pupil size is set from 10 to 45 Mm. The basic processing follows \cite{ZGMZ2011,ZGMZ2012}.

Time--distance diagrams are computed by selecting a source location, rewriting the observed surface velocity signal [$v$] in polar coordinates relative to the source [$ v(r, \theta, t)$] and then using the azimuthal transformation 
\begin{equation}
V_m(r, t)=\int^\beta_{\alpha} v(r, \theta, t) {\mathrm e}^{-{\mathrm i} m \theta} \diff \theta, \label{equ:td}
\end{equation}
where integration is normally performed over the whole circle (i.e. $\alpha=0$, $\beta=2\pi$). Here, we use $m=0,$  and the integration limits  are chosen to represent an arc where the signal is best observed. Acoustic wave-packets of a sufficiently strong amplitude relative to the noise are seen as a time--distance ridge that largely follows a theoretical time--distance curve.

The holography method works by using Green's function [$G_+ (|\vect{r}-\vect{r}'|, t-t'),$] which 
prescribes the acoustic wave propagation from a point source, to essentially ``backtrack'' the observed surface 
signal [$\psi(\vect{r}, t)$]. This allows us to reconstruct egression images showing the subsurface acoustic sources and sinks:
\be
H_+({\vect{r}, z, t})=\int \diff t' \int_{r'=a}^{r'=b}  \int_{\theta=0}^{\theta=2\pi}  G_+ (r', t-t') \psi(r', \theta, t') \diff^2\vect{r}' ,
\label{equ:holo1}
\ee
where $r'$ and $\theta$ are the polar coordinates describing the vector $|\vect{r}-\vect{r}'|$ at the solar surface. Thus, $a, b$ define the dimensions of the holographic pupil. Integration over $\theta$ normally takes place over the whole range $[0, 2\pi]$, while here, in order to quantify and compare the helioseismic directional information, we use the interval of the form $[\theta_0-\delta\theta, \theta_0+\delta\theta,]$ varying $\theta_0$ and taking $\delta\theta=\pi/2$.
 The holographic method is generally more sensitive, but vulnerable to noise, variations from the assumed model and is susceptible to an increased possibility of false detections, which is normally handled by analysing the statistical significance of the detected signal \cite{DL2005,MZZ2011}.

\begin{figure}
\begin{center}
\includegraphics[width=5.25cm]{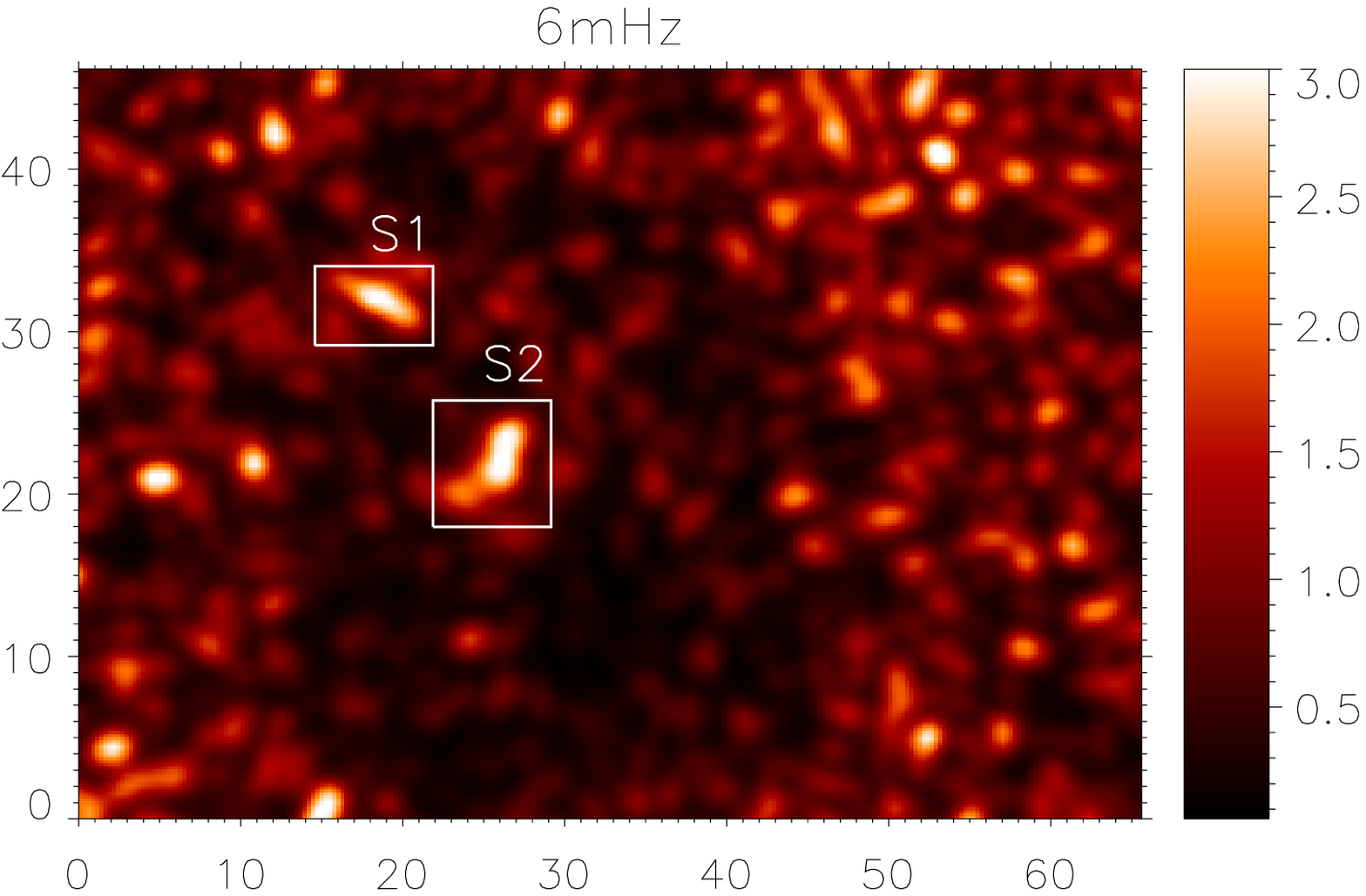} 
\includegraphics[width=5.25cm]{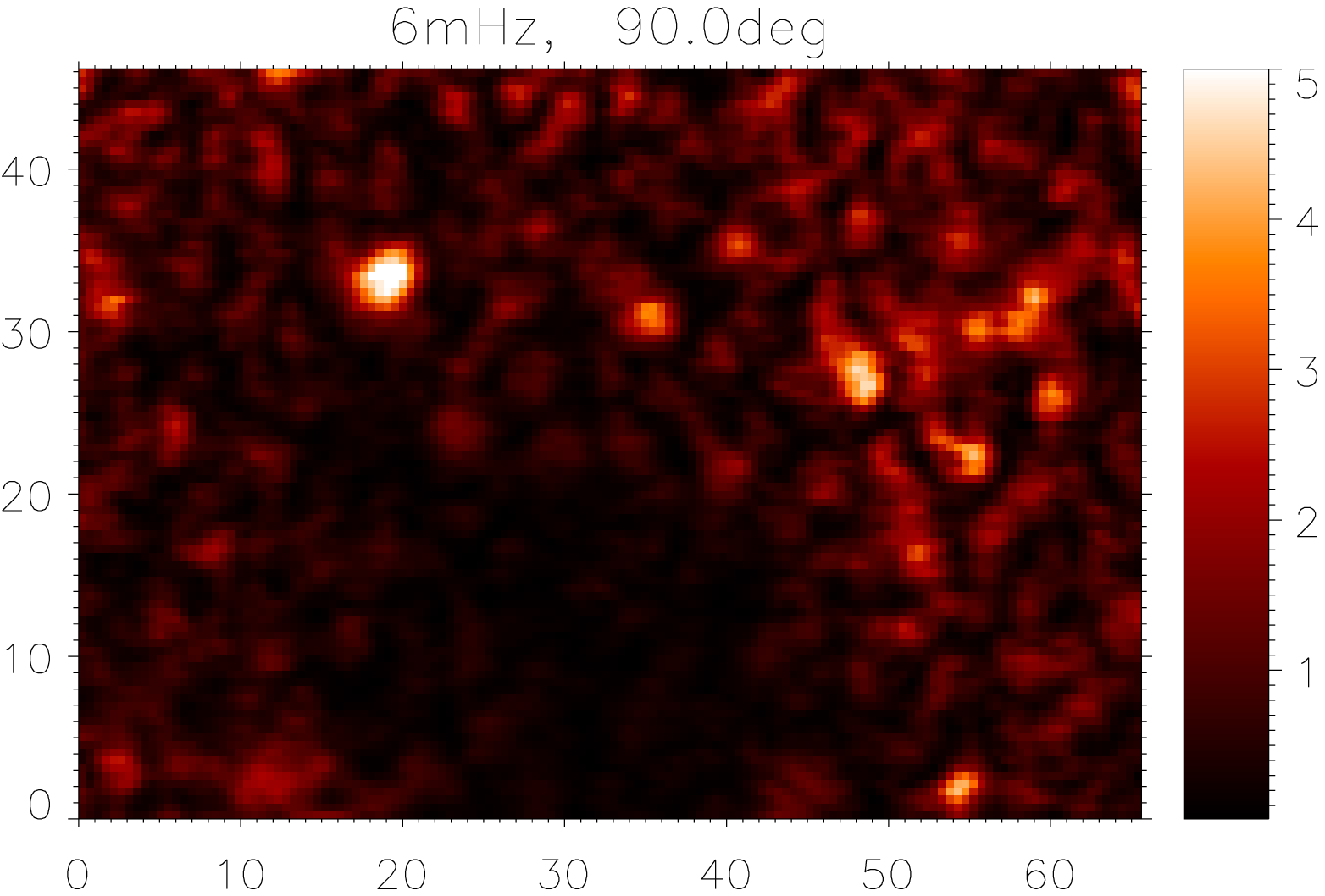}
\includegraphics[width=5.25cm]{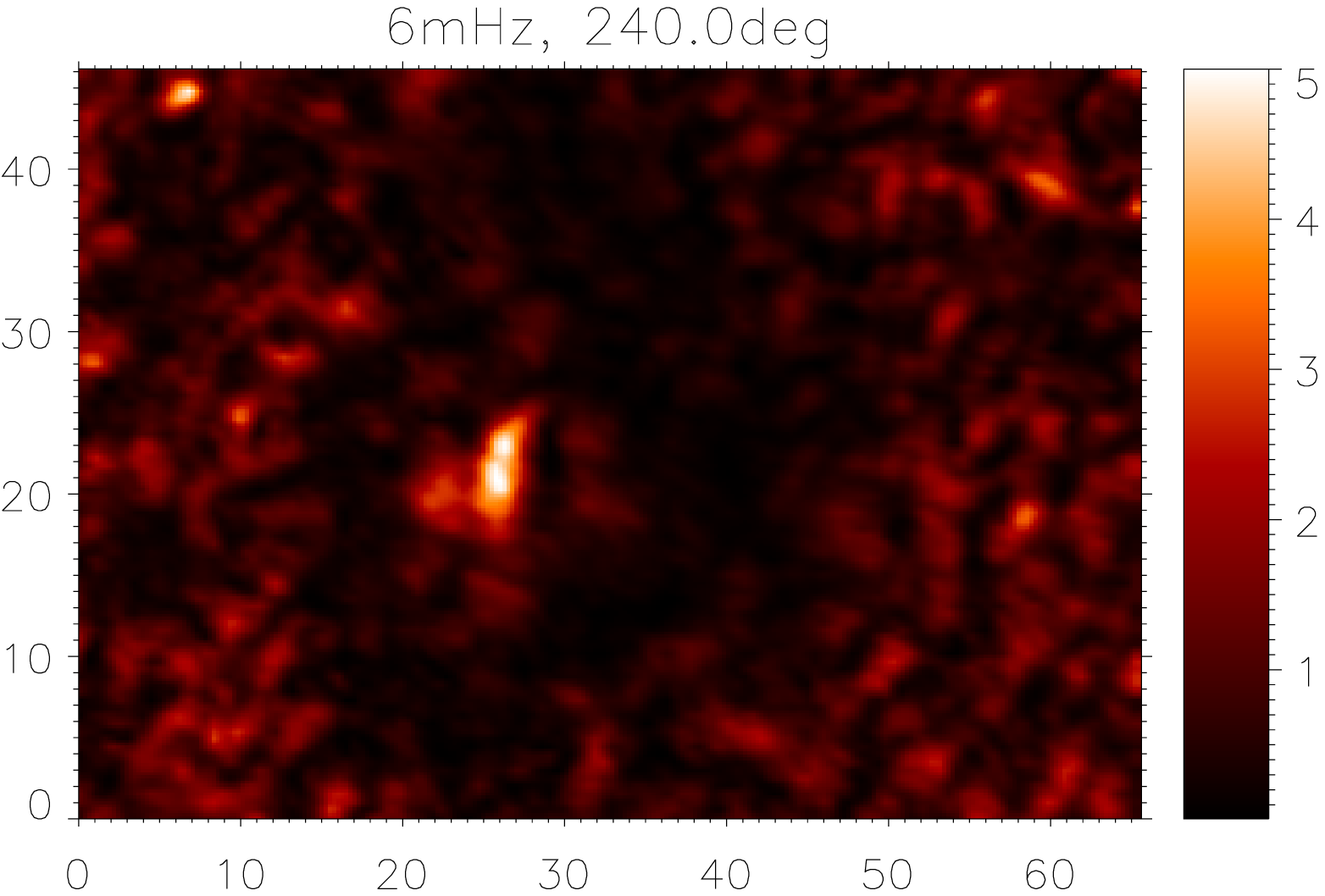}
\end{center}
\caption{ \label{fig:2tds} Egression power snapshot measured at 6mHz taken around 22:35:07\,UT is presented on the left. The integration regions used in deriving statistics for Figure \ref{fig:rms} are presented as white boxes encompassing each source.
This is followed by two directional egression power snapshots with $\theta_0$ measured from $x$-axis counterclockwise given in the title, and $\delta \theta = \pi / 2.$ See Section \ref{sec:method} for details.
All axes are in Mm, field of view is the same as in context images presented in Figure \ref{fig:2sources}.
}
\end{figure}

\begin{figure}
\begin{center}
\includegraphics[width=15cm, height=6.5cm]{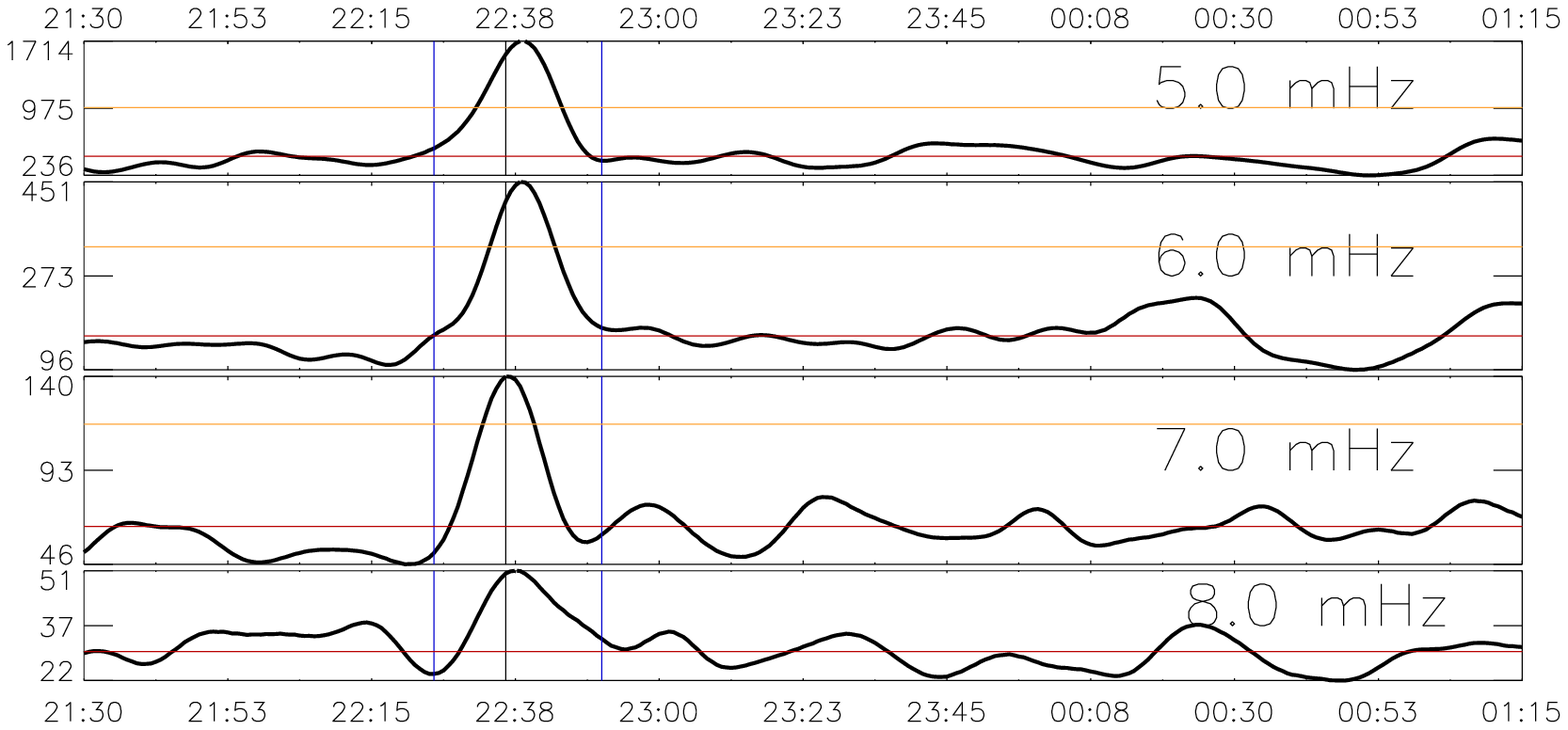}
\includegraphics[width=15cm, height=6.5cm]{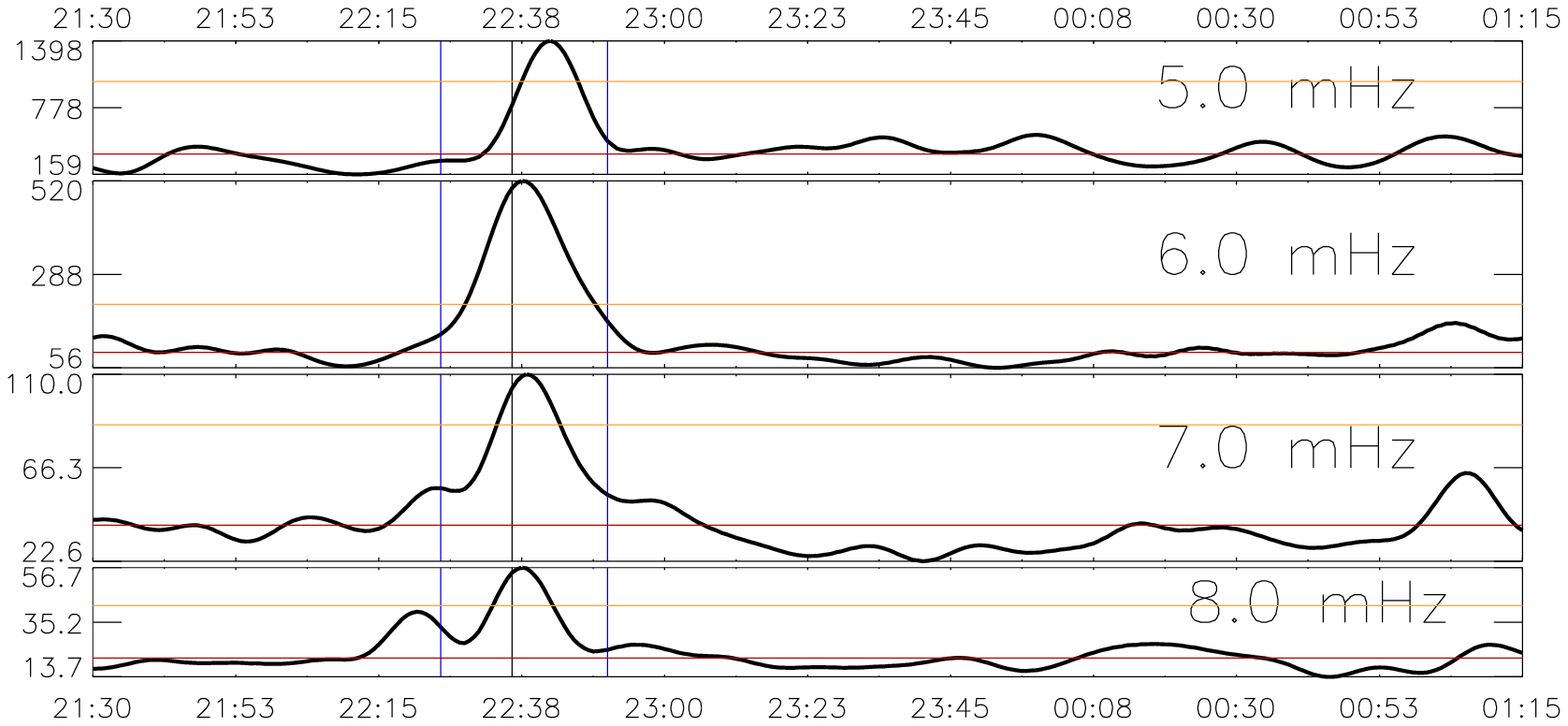}
\end{center}
\caption{\label{fig:rms} Egression rms measured at different frequencies for Source 1 {\em (top)} and Source 2 {\em (bottom)} as indicated in Figure \ref{fig:2tds}. Time is along the $x$-axis, averaged egression power in ${\rm m^2\,s^{-2}}$ is plotted along $y$-axis. Blue vertical lines represent GOES flare start and end times, red horizontal lines show the mean integrated rms measured away from the flare peak, with yellow horizontal line representing the mean plus $5\sigma$ variation significance level.}
\end{figure}

\section{Results}
The context images showing the NOAA 11283 sunspots and magnetic configuration are presented Figure \ref{fig:2sources}. Egression power snapshot computed at 6mHz is presented in left panel of Figure \ref{fig:2tds}. It shows two clear and compact acoustic sources, northern Source 1 (S1) and southern Source 2 (S2). We have found directional time-distance ridges for the two sources which are presented at the bottom right of Figure \ref{fig:2sources} with pointing information shown as orange arcs on top row HMI images. No ridges were found in other directions. To verify the observed acoustic signatures we carry out the rms analysis of the egression power, which is presented in Figure \ref{fig:rms}, showing statistical significance of the measured seismic signal, particularly at 5-6mHz range, for both sources.
Directional holography results are presented in two rightmost panels of Figure \ref{fig:2tds}, showing a clear separation of the sources depending on the direction. The angles are in good general agreement with time-distance measurements, apparently supporting the derived preferred direction of surface wave-front propagation. Interestingly, as can be seen from Figure \ref{fig:2sources}, for Source 2 this direction includes the strongly magnetised plasma of the main sunspot, suggesting that in terms of wave-front anisotropy a mechanism different from magnetic field suppression of acoustic wave is at play. 

Further work is needed in understanding directional results, combined with analysis of RHESSI and SDO AIA observations to get better understanding of the event.
\section*{References}
\bibliography{sdo5}



\end{document}